\RequirePackage{fix-cm}

\documentclass[conference]{IEEEtran}

\usepackage{hyperref}	
\usepackage{setspace}

\ifCLASSINFOpdf
\else
\fi

\usepackage{multirow}
\usepackage{lmodern}
\usepackage{graphicx}
\usepackage{balance}

\usepackage{amsmath}

\usepackage[ruled,lined,boxed,commentsnumbered,linesnumbered]{algorithm2e}
\usepackage{amssymb}

\usepackage{color}
\usepackage[dvipsnames]{xcolor}
\usepackage{tikz,pgfplots}
\usetikzlibrary{shapes,snakes}
\pgfplotsset{compat=1.12}

\usepackage{url}
\usepackage{soul}

\newcommand{\arat}{\alpha}

\IEEEoverridecommandlockouts 
\begin{document}

\title{Efficiently Clustering Very Large Attributed Graphs\thanks{This work has been published in ASONAM 2017~\cite{baroni2017efficiently}. This version includes an appendix with validation of our attribute model and distance function, omitted in~\cite{baroni2017efficiently} for lack of space. Please refer to the published version.}}

\author{
\IEEEauthorblockN{Alessandro Baroni}
\IEEEauthorblockA{University of Pisa, Italy\\
baroni@di.unipi.it }
\and
\IEEEauthorblockN{Alessio Conte}
\IEEEauthorblockA{University of Pisa, Italy\\
conte@di.unipi.it }
\and
\IEEEauthorblockN{Maurizio Patrignani}
\IEEEauthorblockA{Roma Tre University, Italy\\
patrigna@dia.uniroma3.it }
\and
\IEEEauthorblockN{Salvatore Ruggieri}
\IEEEauthorblockA{University of Pisa and\\ ISTI-CNR, Italy\\
ruggieri@di.unipi.it }
}

\maketitle

\begin{abstract}
Attributed graphs model real networks by enriching their nodes with attributes accounting for properties. Several techniques have been proposed for partitioning these graphs into clusters that are homogeneous with respect to both semantic attributes and to the structure of the graph. However, time and space complexities of state of the art algorithms limit their scalability to medium-sized graphs. 
We propose \textbf{SToC} (for \textit{S}emantic-\textit{To}pological \textit{C}lustering), a fast and scalable algorithm for partitioning large attributed graphs.
The approach is robust, being compatible both with categorical and with quantitative attributes, and it is tailorable, allowing the user to weight the semantic and topological components. Further, the approach does not require the user to guess in advance the number of clusters. 
\textbf{SToC} relies on well known approximation techniques such as {bottom-k} sketches, traditional graph-theoretic concepts, and a new perspective on the composition of heterogeneous distance measures. Experimental results demonstrate its ability to efficiently compute high-quality partitions of large scale attributed graphs.
\end{abstract}

\section{Introduction}
Several approaches in the literature aim at partitioning a graph into communities that share some sets of properties (see~\cite{coscia2011classification} for a survey).
Most criteria for defining communities in networks are based on topology and focus on specific features of the network structure, such as the presence of dense subgraphs or other edge-driven characteristics.
However, real world graphs such as the World Wide Web and Social Networks are more than just their topology. A formal representation that is gaining popularity for describing such networks is the \emph{attributed graph}~\cite{bothorel2015clustering,diestel2005graph,gunnemann2011db,zhou2010clustering}.%

An attributed graph is a graph where each node is assigned values on a specified set of attributes. Attribute domains may be either categorical (e.g., sex) or quantitative (e.g., age).
Clustering attributed graphs consists in partitioning them into disjoint communities of nodes that are both well connected and similar with respect to their attributes.
State of the art algorithms for clustering attributed graphs have several  limitations~\cite{zhou2010clustering}: 


they are too slow to be compatible with big-data scenarios, both in terms of asymptotic time complexity and in terms of running times; they use data-structures that need to be completely rebuilt if the input graph changes; and they ask the user to specify the number of clusters to be produced. Moreover, they only work with categorical attributes, forcing the user to discretize the domains of quantitative attributes, leading to a loss of  information in distance measures.

We offer a new perspective on the composition of heterogeneous distance measures. Based on this, we present a distance-based clustering algorithm for attributed graphs that allows the user to tune the relative importance of the semantic and structural information of the input data and only requires to specify as input qualitative parameters of the desired partition rather than quantitative ones (such as the number of clusters). This approach is so effective that can be used to directly produce a set of similar nodes that form a community with a specified input node without having to cluster the whole graph first.
We rely on state-of-the-art approximation techniques, such as {bottom-k} sketches for approximating similarity between sets~\cite{cohen2007summarizing} and the Hoeffding bound (see~\cite{crescenzi2011comparison}) to maintain high performance while keeping the precision under control.
Regarding efficiency, our approach has an expected time complexity of $O(m \log n)$, where $n$ and $m$ are the number of nodes and edges in the graph, respectively. This performance is achieved via the adoption of a distance function that can be efficiently computed in sublinear time. Experimental results demonstrate the ability of our algorithm to produce high-quality partitions of large attributed graphs.

The paper is structured as follows. Section~\ref{sec:related} describes related work. Section~\ref{sec:contributions} summarizes our contributions. Section~\ref{sec:problem} formally states the addressed problem. Section~\ref{sec:distance} introduces the notion of distance, essential for cluster definition. The algorithm and its data structures are described in Section~\ref{se:algorithm}. Section~\ref{sec:tuning} describes the tuning phase which selects suitable parameter values for the input graph according to the user's needs. The results of our experimentation are discussed in Section~\ref{sec:experiments}. Finally, Appendix~\ref{apx:model} and~\ref{apx:sctoc} contain a validation of our attribute model and combined distance function.

\section{Related work}\label{sec:related}

Graph clustering, also known as community discovery, is an active research area (see~e.g.,~the survey \cite{coscia2011classification}). While the overwhelming majority of the approaches assign nodes to clusters based on topological information only, recent works address the problem of clustering graphs with semantic information attached to nodes or edges (in this paper, we restrict to node-attributed graphs). In fact, topological-only or semantic-only clustering approaches are not as effective as approaches that exploit both sources of information \cite{ding2011community}. A survey of the area \cite{bothorel2015clustering} categorizes the following classes for node-attributed graphs (here we recall only key references and uncovered recent papers).
\emph{Reduction-based} approaches translate the semantic information into the structure of the graph (for example adding weights to its edges) or vice versa (for example encoding topological information into new attributes), then perform a traditional clustering on the obtained data. They are efficient, but the quality of the clustering is poor. 
\emph{Walk-based} algorithms augment the graph with dummy nodes representing categorical only attribute values and estimate node distance through a neighborhood random walk~\cite{zhou2010clustering}. The more attributes two nodes shares the more paths connect them the more the nodes are considered close. A traditional clustering algorithm based on these distances produces the output. 
\emph{Model-based} approaches statistically infer a model of the clustered attributed graphs, assuming they are generated accordingly to some parametric distribution. The \textbf{BAGC} algorithm \cite{xu2012model} adopts a Bayesian approach to infer the parameters that best fit the input graph, but requires the number of clusters to be known in advance and does not handle quantitative attributes. The approach has been recently generalized to weighted attributed graphs in \cite{xu2014gbagc}. The resulting \textbf{GBAGC} algorithm is the best performing of the state-of-the-art (in Section~\ref{sec:experiments} we will mainly compare to this work). A further model-based algorithm is \textbf{CESNA}~\cite{yang2013community}, which addresses the different problem of discovering overlapping communities. 
\emph{Projection-based} approaches focus on the reduction of the attribute set, omitting attributes are irrelevant for some clusters. A recent work along this line is \cite{sanchez2015efficient}. 
Finally, there are some approaches devoted to variants of attributed graph clustering, such as: 
\textbf{I-Louvain} \cite{combe2015louvain}, which extends topological clustering to maximize both modularity and a newly introduced measure called `inertia'; 
\textbf{PICS} \cite{akoglu2012pics}, which addresses a form of co-clustering for attributed graphs by using a matrix compression-approach; 
\textbf{FocusCO} \cite{perozzi2014focused} and \textbf{CGMA} \cite{cao2016user}, which start from user-preferred clusters; 
\textbf{M-CRAG} \cite{Guy2009PersonalizedRecommendation}, which generates multiple non-redundant clustering for exploratory analysis; 
\textbf{CAMIR} \cite{papadopoulos2015clustering}, which considers multi-graphs.

Overall, state-of-the-art approaches to partition attributed graphs are affected by several limitations, the first of which is efficiency. 
Although, the algorithm in~\cite{zhou2010clustering} does not provide exact bounds, our analysis assessed an $\Omega(n^2)$ time and space complexity, which restricts its usability to networks with thousands of nodes. The algorithm in~\cite{xu2014gbagc} aims at overcoming these performance issues, and does actually run faster in practice. However, as we show in Section~\ref{sec:experiments}, its time and space performances heavily rely on assuming a small number of clusters. 
Second, similarity between elements is usually defined with exact matches on categorical attributes, so that similarity among quantitative attributes is not preserved.
Further, data-structures are not maintainable, so that after a change in the input graph they will have to be fully recomputed.
Finally, most of the approaches require as input the number of clusters that have to be generated~\cite{kanungo2002efficient,von2007tutorial,zhou2010clustering,xu2014gbagc}.
In many applications it is unclear how to choose this value or how to evaluate the correctness of the choice, so that the user is often forced to repeatedly launching the algorithm with tentative values.

\section{Contributions}\label{sec:contributions}

We propose an approach to partition attributed graphs that aims at overcoming the limitations of the state of the art discussed in Section~\ref{sec:related}. 
Namely: (i) We propose a flexible concept of distance that can be efficiently computed and that is both tailorable (allowing the user to tune the relative importance of the semantic and structural information) and robust (accounting for both categorical and quantitative attributes). Further, our structures can be maintained when entities are added or removed without re-indexing the whole dataset.
(ii) We present the \textbf{SToC} algorithm to compute non-overlapping communities. 
\textbf{SToC} allows for a declarative specification of the desired clustering, i.e.,~the user has to provide the sensitivity with which two nodes are considered close rather than forecast the number of clusters in the output.
(iii) We describe an experimental comparison with state-of-the-art approaches showing that \textbf{SToC} uses less time/space resources and produces better quality partitions. In particular, in addition to good quality metrics for the obtained clusters, we observe that \textbf{SToC} tends to generate clusters of homogeneous size, while most partitioning algorithms tend to produce a giant cluster and some smaller ones.

\section{Problem statement}\label{sec:problem}

Intuitively, attributed graphs are an extension of the structural notion of graphs to include attribute values for every node. 
Formally, an \emph{attributed graph} $G(V,E,F)$ consists of a set $V = \{v_1, \ldots,v_n\}$ of \emph{nodes}, a set $E=\{e_1,\ldots,e_m\} \subseteq V \times V$ of \emph{edges},  
and a set of mappings $F=\{f_1,\ldots,f_A\}$ such that, for $i \in [1..A]$, $f_i: V \rightarrow dom(a_i)$ assigns to a node $v$ the value $f_i(v)$ of attribute $a_i$, where $dom(a_i)$ is the domain of attribute $a_i$. Notice that the definition is stated for directed graphs, but it readily applies to undirected ones as well. 
A \emph{distance function} $d: V \times V \rightarrow \mathbb{R}_{\geq 0}$ quantifies the dissimilarity between two nodes through a non-negative real value, where $d(v_1, v_2) = d(v_2, v_1)$, and $d(v_1, v_2) = 0$ iff $v_1 = v_2$. Given a threshold $\tau$, the \emph{ball centered at node~$v$}, denoted $B_{d}(v, \tau)$, consists of all nodes at distance at most~$\tau$:
$B_{d}(v, \tau) = \{ v' \in V \ |\ d(v, v') \leq \tau \}.$
We distinguish between \emph{topological distances}, that are only based on structural properties of the graph $G(V, E)$, \emph{semantic distances}, that are only based on node attribute values $F$, and \emph{multi-objective distances}, that are based on both~\cite{deza2009encyclopedia}.
Using on distance functions, a \emph{cluster} can be defined by considering nodes that are within a maximum distance $\tau$ from a given node. 
Namely, for a distance function $d()$ and a threshold $\tau$, a \emph{$\tau$-close cluster} $C$ is a subset of the nodes in $V$ such that there exists a node $v \in C$ such that for all $v' \in C$, $d(v, v') \leq \tau$. The node $v$ is called a \emph{centroid}.
Observe that $C \subseteq B_d(v, \tau)$ is required but the inclusion may be strict. In fact, a node $v' \in B_d(v, \tau)$ could belong to another $\tau$-close cluster $C'$ due to a lower distance from the centroid of $C'$. 

A \emph{$\tau$-close clustering} of an attributed graph is a partition of its nodes into $\tau$-close clusters $C_1, \ldots, C_k$. 
Notice that this definition requires that clusters are disjoint, i.e.,~non-overlapping.

\newcommand{\colNA}{gray!50}
\newcommand{\colNB}{white}

\begin{figure*}
 \centering
 \scalebox{0.9}{ 
 \parbox{.55\textwidth}{
 \noindent
 \scalebox{0.7}{
 	\begin{tikzpicture}
        \begin{axis}[
        hide axis, scale only axis, height=0pt, width=0pt, 
        colormap/jet,
        colorbar sampled,
        colorbar horizontal,
        point meta min = 0,
        point meta max = 1,
        colorbar style = {
            samples = 10,
            height = 0.2cm,
            width = 10cm,
            xtick style={draw=none},
            xticklabel style = {
                text width = 2.5em,
                align = center,
                /pgf/number format/.cd,
                fixed,
                fixed zerofill,
                precision = 1,
                /tikz/.cd
            }
        }
        ]
        \addplot [draw=none] coordinates {(0,0)};
        \end{axis}
        \def\len{5mm}
        \definecolor{leftcolor}{RGB}{191, 191, 191}
        \definecolor{rightcolor}{RGB}{102, 102, 102}
        \foreach \i/\j in {south east/a,east/b,north east/c,north west/e,west/f,south west/g}
        {\coordinate (\j) at (current colorbar axis.\i);}
    \end{tikzpicture}
    } \scalebox{0.85}{
	\begin{tikzpicture}[baseline=(B.base), shorten >=1pt, auto, thick,node distance=1cm,
	main node/.style={circle,draw,font=\sffamily\small\bfseries,minimum  size=0.1cm},
	male node/.style={diamond,draw,font=\sffamily\footnotesize\bfseries,minimum size=0.05cm,fill=\colNA},
	female node/.style={regular polygon,regular polygon sides=9,draw,font=\sffamily\footnotesize\bfseries,
	                    minimum  size=0.05cm,fill=\colNB}]
	    \coordinate (B) at (0,0.8);
	    \coordinate (XY0) at (-1,-1);
        \coordinate (X1) at (7,-1);
        \coordinate (Y1) at (-1,3);
        \draw [ -> ] (XY0) -- (X1) node [right] {x};
        \draw [ -> ] (XY0) -- (Y1) node [right] {y};
        \coordinate (Node0) at (0,1);
        \coordinate (Node1) at (0,0);
        \coordinate (Node2) at (1,1);
        \coordinate (Node3) at (2,0);
        \coordinate (Node4) at (4,0);
        \coordinate (Node5) at (5.5,1);
        \coordinate (Node6) at (6,0);
        \coordinate (Node7) at (7,0.9);
        \node[main node,fill=\colNA] (0) at (Node0) {$v_0$};        
        \node[main node,fill=\colNA] (1) at (Node1) {$v_1$};
        \node[main node,fill=\colNA] (2) at (Node2) {$v_2$};
        \node[main node,fill=\colNB] (3) at (Node3) {$v_3$};
        \node[main node,fill=\colNB] (4) at (Node4) {$v_4$};
        \node[main node,fill=\colNB] (5) at (Node5) {$v_5$};
        \node[main node,fill=\colNB] (6) at (Node6) {$v_6$};
        \node[main node,fill=\colNA] (7) at (Node7) {$v_7$};
		\path [line width=1mm] (0) edge [cyan] node {} (1);
		\path [line width=1mm] (0) edge [NavyBlue] node {} (2);
		\draw [color=Orange,line width=1mm] (0) to [out=50, in=130] (7);
		\path [line width=1mm] (1) edge [NavyBlue] node {} (2);
        \path [line width=1mm] (1) edge [Goldenrod] node {} (3);
        \path [line width=1mm] (3) edge [Goldenrod] node {} (4);
        \path [line width=1mm] (4) edge [cyan] node {} (5);
        \path [line width=1mm] (4) edge [cyan] node {} (6);
        \path [line width=1mm] (5) edge [blue] node {} (6);
		\path [line width=1mm] (5) edge [cyan] node {} (7);
		\path [line width=1mm] (6) edge [cyan] node {} (7);
	\end{tikzpicture}
	}} \scalebox{0.85}{
	\begin{tabular}{|l|l|l|l|}
			\hline
			ID&sex&x&y\\		\hline	\hline
			0&1&0&0.1\\
			1&1&0&0\\
			2&1&0.1&0.1\\
			3&0&0.2&0\\
			4&0&0.4&0\\
			5&0&0.55&0.1\\
			6&0&0.6&0\\
			7&1&0.7&0.09\\\hline
	\end{tabular}
	\begin{tabular}{|l|l|} 
		\hline
		$d_S(v_0, v_i)$&$d_T(v_0, v_i)$\\ \hline \hline
		0&0\\
		0.00471&0.4\\
		0.0066&0.25\\
		0.3427&0.6\\
		0.3521&0.86\\
		0.3596&1\\
		0.3616&1\\
		0.3327&0.86\\ \hline
	\end{tabular}}
	}
	\caption{A sample attributed graph. Attributes include: sex (categorical) and spatial coordinate $x$, $y$ (quantitative). Edges are colored by topological distance between the connected nodes. Topological distance $d_T()$ considers $1$-neighborhoods.}
	\label{fig:smallGraph}
\end{figure*}
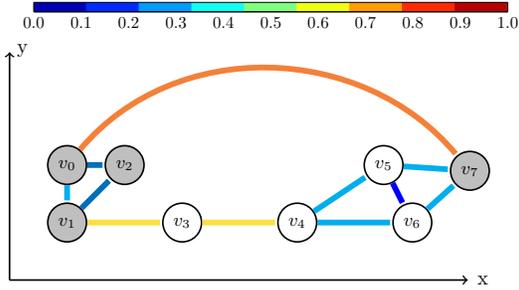

\section{A multi-objective distance}\label{sec:distance}

In this section we introduce a multi-objective distance function that will be computable in sublinear time (Section~\ref{subsec:stoQuery}) and that allows for tuning the semantic and topological components (Section~\ref{sec:tuning}).
Regarding semantic distance, we assume that attributes can be either categorical or quantitative. For clarity, we assume that attributes $1, \ldots, Q$ are quantitative, and attributes $Q+1, \ldots, A$ are categorical.
Thus, the attribute values $(f_1(v), \ldots, f_A(v))$ of a node $v$, boil down to a relational tuple, which we denote by~$\mathbf{t}_v$.
Semantic distance $d_S()$ can then be defined as: 
\begin{equation}\label{eq:hybridDistance}
d_S(v_1, v_2)= f( \mathbf{t}_{v_1}, \mathbf{t}_{v_2})
\end{equation}
where $f()$ is any distance function over relational tuples (see e.g.,~\cite{HanKamber11}). 
In our implementation and in experiments, we first normalize quantitative attributes in the $[0, 1]$ interval using min-max normalization~\cite{mohamad2013standardization}. Then, we
adopt the following distance:
\begin{equation}
\tiny
d_s(v_1,v_2) = \frac{\sqrt{\sum_{i=1}^{Q} 
((f_i(v_1) - f_i(v_2))^2} \cdot \sqrt{Q} + \sum_{i=Q+1}^{A} J(f_i(v_1),f_i(v_2))}{A}
\label{equ:semanticDistance}
\end{equation}
Quantitative attributes are compared using Euclidean distance, and categorical attributes are compared using Jaccard distance ($J(X, Y) = 1 - \frac{|X \cap Y|}{|X \cup Y|}$).
The scaling factor $\sqrt{Q}$ balances the contribution of every quantitative attribute in the range $[0,1]$. The overall distance function ranges in $[0, 1]$.

Regarding topological distance, we assume it is defined as a distance of the node neighborhoods.
Let us recall the notion of $l$-neighborhood from~\cite{gunnemann2011db}: 
the $l$-neighborhood $N_l(v)$ of $v$ is the set of nodes reachable from $v$ with a path of length at most~$l$. $N(v)$ is a shorthand for $N_1(v)$, namely the nodes linked by $v$.

Topological distance $d_T()$ can then be defined as: 
\begin{equation}\label{eq:structuralDistance}
d_T(v_1, v_2)= g( N_l(v_1), N_l(v_2))
\end{equation}
where $g()$ is any distance function over sets of nodes, e.g.,~Dice, Jaccard, Tanimoto \cite{deza2009encyclopedia}. In particular, we adopt the Jaccard distance.
Fig.~\ref{fig:smallGraph} shows an example on distances. Fix node $v_0$, and consider  distances (semantic and topological) between $v_0$ and the other nodes in the graph. For topological distance (with $l=1$), e.g.,~we have $d_T(v_0,v_2) = J(\{v_0,v_1,v_2,v_7\}, \{v_0,v_1,v_2\}) =  1-\frac{|\{v_0,v_1,v_2\}|}{|\{v_0,v_1,v_2,v_7\}|} = 0.25$.
Thus, $v_0$ is closer to $v_2$ than to $v_1$, since $d_T(v_0,v_1) = 0.4$. For semantic distance (Equ.~\ref{equ:semanticDistance}), instead, it turns out that $v_0$ is closer to $v_1$ than to $v_2$. In fact, all three of them have the same sex attribute, but $v_1$ is spatially closer to $v_0$ than to $v_2$.

Finally, semantic and topological distance can be combined into a multi-objective distance $d_{ST}()$ as follows \cite{bothorel2015clustering}:
\begin{equation}\label{eq:combinedDistance}
d_{ST}(v_1, v_2) = h( d_S(v_1, v_2) ,  d_T(v_1, v_2) )
\end{equation}

where $h: \mathbb{R}_{\geq 0} \times \mathbb{R}_{\geq 0} \rightarrow \mathbb{R}_{\geq 0}$ is such that $h(x, y) = 0$ iff $x = y = 0$. This and the assumptions that $d_S()$ and $d_T()$ are distances imply that $d_{ST}()$ is a distance.
\cite{combe2012combining,villa2013carte,zhou2010clustering} set $h(x, y) = x + y$ as the sum of semantic and topological distances.  If $x \gg y$ then $h(x, y) \approx x$, so the largest distance weights more. However, if $x \approx y$ then $h(x, y) \approx 2x$, which doubles the contribution of the equal semantic and topological distances. In this paper, we consider instead $h(x, y) = max(x,y)$. If $x \gg y$ then $h(x, y) \approx x \approx y$, as before. However, when $x \approx y$ then $h(x, y) \approx x$, which agrees with the distance of each component.

\section{The \textbf{SToC} Algorithm}\label{se:algorithm}

For a given distance threshold $\tau$, \textbf{SToC} iteratively extracts $\tau$-close clusters from the attributed graph starting from random seeds. This is a common approach in many clustering algorithms~\cite{bothorel2015clustering,coscia2011classification}. Nodes assigned to a cluster are not considered in the subsequent iterations, thus the clusters in output are not overlapping. The algorithm proceeds until all nodes have been assigned to a cluster.
This section details the \textbf{SToC} algorithm. We will proceed bottom-up, first describing the \textbf{STo-Query} procedure which computes a $\tau$-close cluster with respect to a given seed $s$, and the data structures that make its computation efficient. Then, we will present the main \textbf{SToC} procedure.

\subsection{The STo-Query Procedure}
\label{subsec:stoQuery}

The \textbf{STo-Query} procedure computes a connected $\tau$-close cluster $C$ for a given seed $s$ and threshold $\tau$. With our definition of $d_{ST}()$, it turns out that $C \subseteq B_{d_{ST}}(s, \tau) = B_{d_{S}}(s, \tau) \cap B_{d_{T}}(s, \tau)$.
We define $C$ as the set of nodes in $B_{d_{ST}}(s, \tau)$ that are connected to $s$. Computing $C$ is then possible through a partial traversal of the graph starting from $s$. This is the approach of the \textbf{STo-Query} procedure detailed in Algorithm~\ref{alg:stoq}.

The efficiency of  \textbf{STo-Query} relies on two data structures, $S$ and $T$, that we adopt for computing semantic and topological distances respectively and, a fortiori, for the test $d_{ST}(s, x) \leq \tau$ at line 7. Recall that $d_{ST}(s, x) = max( d_{S}(s, x), d_{T}(s, x) )$.
Semantic distance is computed by directly applying (Equ.~\ref{equ:semanticDistance}). We store in a dictionary $S$ the mapping of nodes to attribute values. Thus, computing $d_{S}(s, x)$ requires $O(A)$ time, where $A$ is the number of attributes.
For topological distance, instead, a na\"ive usage of  (Equ.~\ref{eq:structuralDistance}) would require to compute online the $l$-neighborhood of nodes. This takes $O(n)$ time for medium-to-large values of $l$, e.g.,~for small-world networks. We overcome this issue by approximating the topological distance with a bounded error, by using \textit{bottom-k} sketch vectors~\cite{cohen2007summarizing}.
A sketch vector is a compressed representation of a set, in our case an $l$-neighborhood, that allows the estimation of functions, in our case topological distance, with bounded error. 
The bottom-$k$ sketch $S(X)$ consists in the first $k$ elements of a set $X$ with respect to a given permutation of the domain of elements in $X$.
The Jaccard distance between $N_l(v_1)$ and $N_l(v_2)$ can be approximated using $S(N_l(v_1))$ and $S(N_l(v_2))$ in their place, with  a precision $\epsilon$ by choosing $k = \frac{\log{n}}{\epsilon^2}$ (see~\cite{crescenzi2011comparison}).  We store in a dictionary $T$ the mappings of nodes $v$ to the sketch vector of $N_l(v)$.
$T$ allows for computing $d_{T}(s, x)$ in $O(\log{n})$ time.

\begin{algorithm}[t]
\SetKwData{G}{G}
\SetKwData{GG}{G(V,E)}
\SetKwData{V}{V} 
\SetKwData{info}{$\mathbf{A}$} 
\SetKwData{thresh}{$\tau$} 
\SetKwData{lsh}{$S$} 
\SetKwData{bbls}{$T$} 
\SetKwData{comms}{C} 
\SetKwData{com}{$C_{s}$} 
\SetKwData{node}{s} 
\SetKwData{cnode}{v} 
\SetKwData{buck}{B} 
\SetKwData{cand}{y} 
\SetKwData{enq}{\textsc{enqueue}} 
\SetKwData{deq}{\textsc{dequeue}} 
\SetKwFunction{dist}{$d_{ST}$}
\DontPrintSemicolon
\SetKwInOut{Input}{Input}
\SetKwInOut{Output}{Output}
\SetKwInOut{Global}{Global}
\Input{$G$, $\tau$, $\node$}
\Output{$C$, a $\tau$-close connected cluster}
\Global{\lsh and \bbls data structures (described in Section~\ref{subsec:stoQuery}, used to compute $d_{ST}$)}
\BlankLine
$Q \leftarrow$ empty queue\;
$C \leftarrow \{s\} $\;
$\enq(Q,s)$\;
\While{$Q \neq \emptyset$}{
    $v \gets \deq(Q)$\;
    \ForEach{$x \in N(v)$}{
        \If{$x\not\in C$ and $d_{ST}(s,x) \leq \tau$}{
            $C \gets C \cup \{x\}$\;
            $\enq(Q,x)$\;
        }
    }
}
\Return $C$;
\BlankLine
\caption{\textbf{STo-Query} algorithm.}
\label{alg:stoq}\vspace{-0.2cm}
\end{algorithm}
\subsection{The SToC Procedure}
The algorithm consists of repeated calls to the \textbf{STo-Query} function on selected seeds. $\tau$-close clusters returned by \textbf{STo-Query} are added to output and removed from the set of active nodes $V'$ in  Algorithm~\ref{alg:clustering} (lines 7--9). We denote by $G[V']$ the subgraph of $G$ with only the nodes in $V'$.
Seeds are chosen randomly among the active nodes through the \texttt{select\_node} function (line 6). This philosophy can be effective in real-world networks (see, e.g.,~\cite{Conte:2016:CCL:2851613.2851816}), and is inherently different from selecting a set of random seeds in the beginning (as in k-means), since it guarantees that each new seed will be at a significant distance from previously chosen ones.
Calls to \textbf{STo-Query} return non-overlapping clusters, and the algorithm terminates when all nodes have been assigned to some cluster, i.e.,~$V' = \emptyset$.

\begin{algorithm}[t]
\SetKwData{G}{G}
\SetKwData{GG}{G(V, E, F)}
\SetKwData{V}{V}
\SetKwData{info}{F} 
\SetKwData{thresh}{$\tau$} 
\SetKwData{lsh}{$S$} 
\SetKwData{bbls}{$T$} 
\SetKwData{clust}{R} 
\SetKwData{parent}{parent} 
\SetKwData{com}{$C$} 
\SetKwData{node}{s} 
\SetKwData{cnode}{v} 
\SetKwData{id}{id} 
\SetKwFunction{dist}{$d_{ST}$}
\SetKwFunction{sel}{select\_node}
\SetKwFunction{query}{STo-Query}
\DontPrintSemicolon
\SetKwInOut{Input}{Input}
\SetKwInOut{Output}{Output}
\Input{\GG attributed graph, \thresh distance threshold}
\Output{\clust, a $\tau$-close clustering of $G$}
\BlankLine
\lsh $\gets$ global dictionary of the semantic vectors\;
\bbls $\gets$ global topological similarity table of \G\;
\clust $\gets \emptyset$\;
$V' \gets V$\;
\While{$V' \neq \emptyset$}{
    \node $\gets \sel{G}$\;
    \com $\gets $ \query{$G$, $\tau$, $\node$} \;
    $V' \gets V' \setminus C$\;
    $G \gets G[V']$\;
    $\clust \gets \clust \cup \{ \com \}$\;
}

\Return \clust\;
\BlankLine
\caption{ \textbf{SToC} algorithm.}
\label{alg:clustering}\vspace{-0.2cm}
\end{algorithm}

\begin{figure*}
    \centering
    \includegraphics[width=.55\columnwidth]{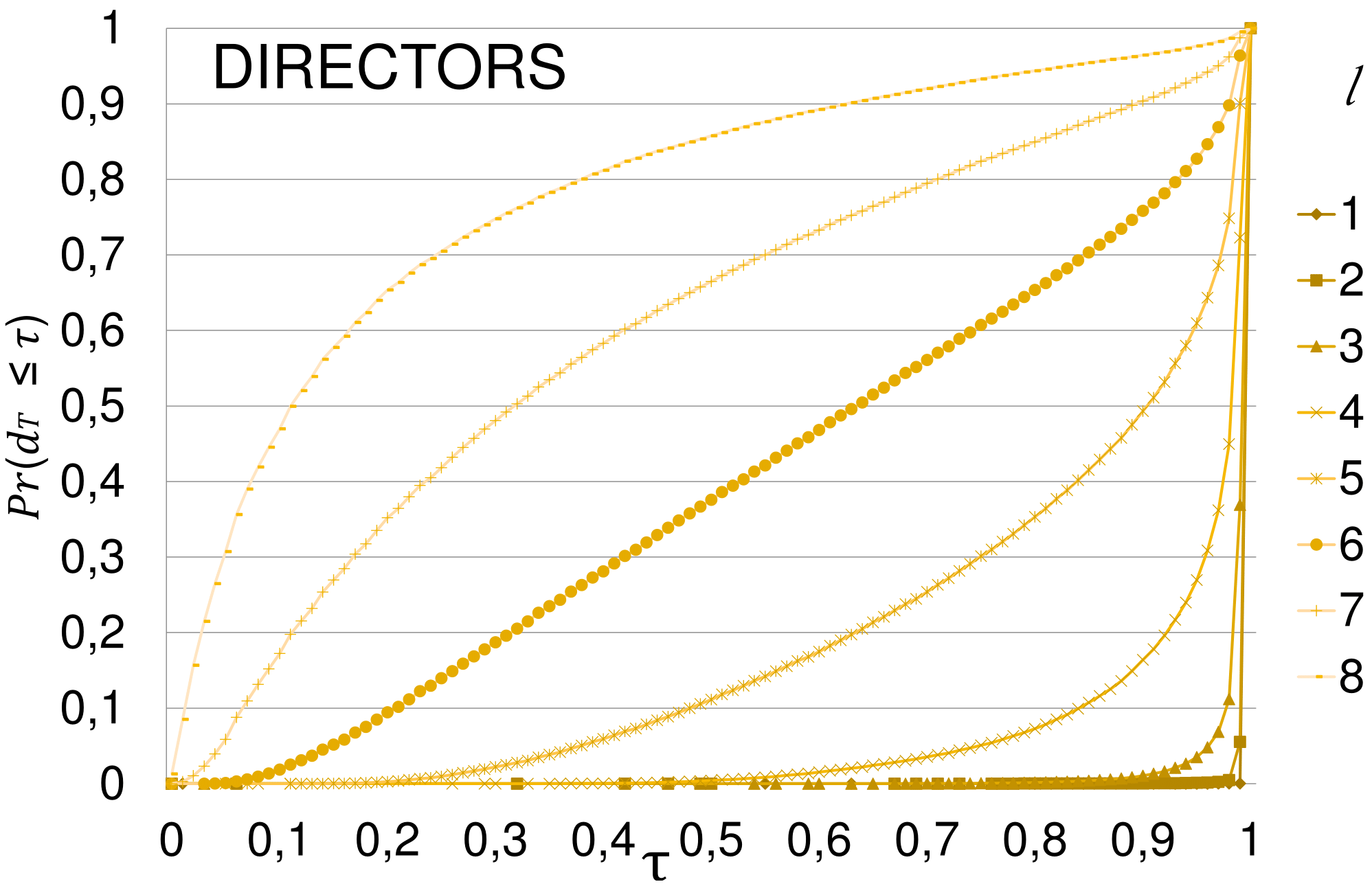}
    \includegraphics[width=.45\columnwidth]{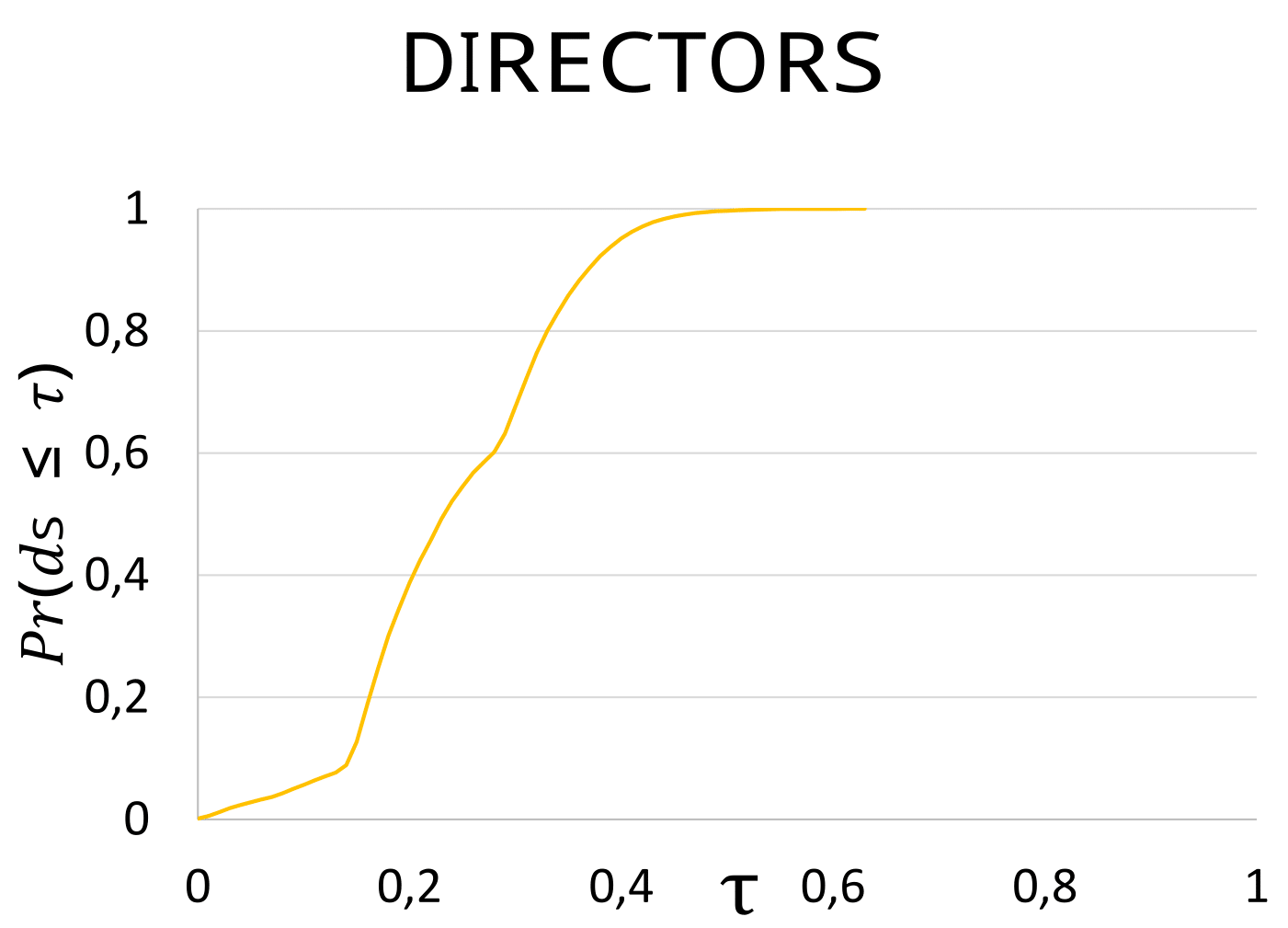}
    \includegraphics[width=.55\columnwidth]{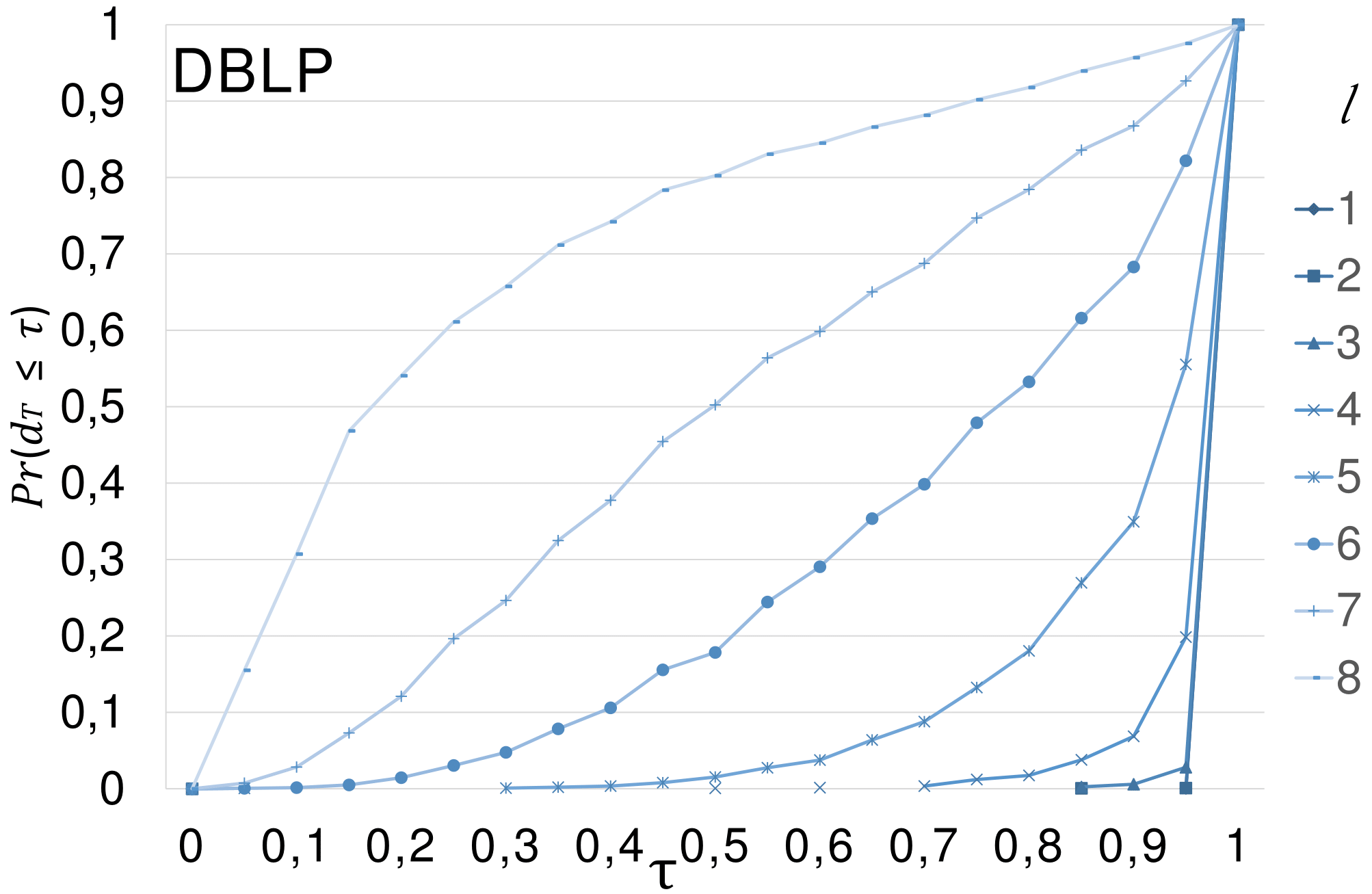}
    \includegraphics[width=.45\columnwidth]{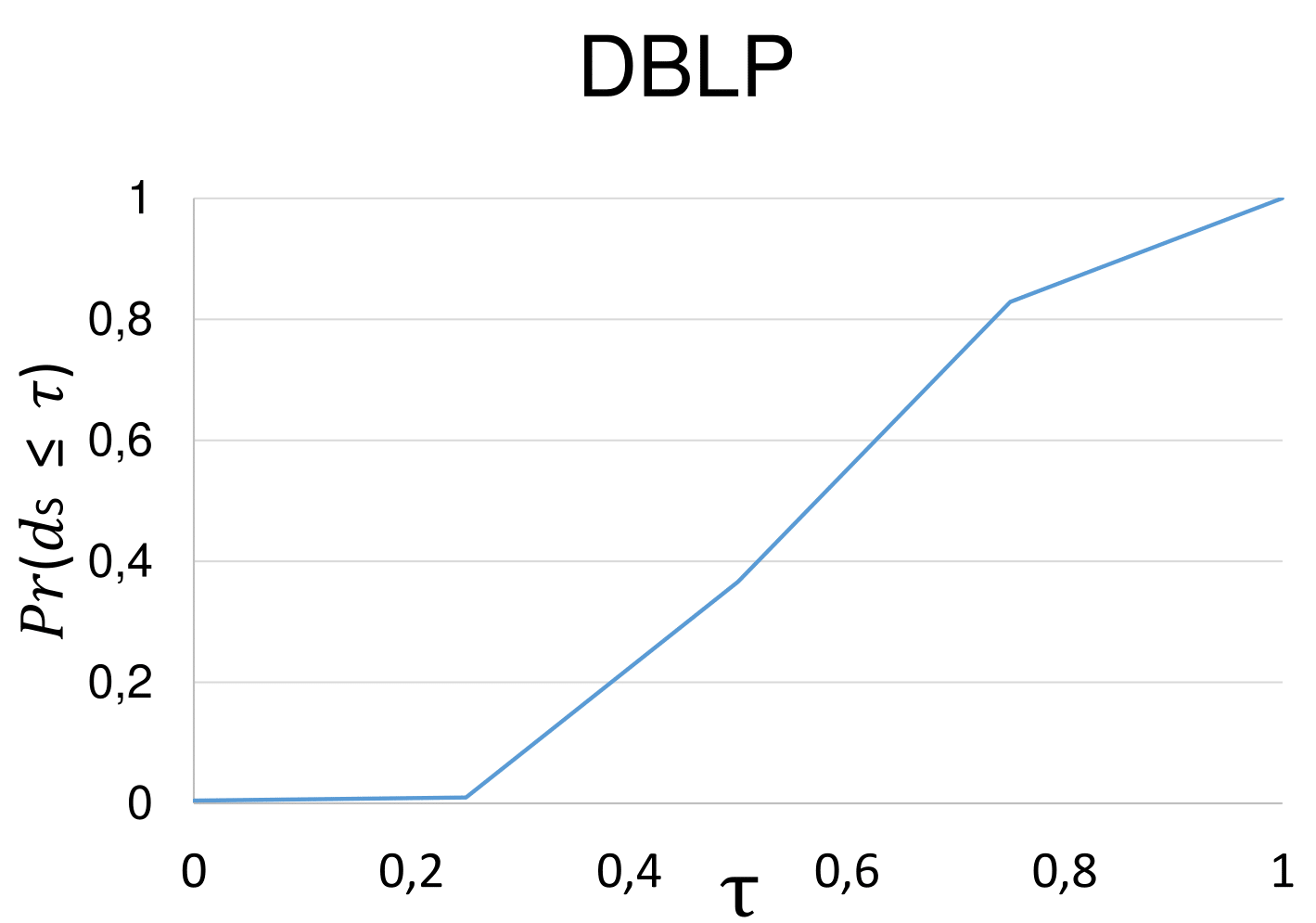}
    \caption{Cumulative distributions of $d_{T}()$, for varying $l$-neighborhoods, and of $d_S()$ for datasets DIRECTORS (left) and DBLP (right).
    \label{fig:ddists}}
\end{figure*}

\subsection{Time and space complexity}\label{sec:cost}

Let us consider time complexity first.
For a seed $s$, \textbf{STo-Query} iterates over nodes in $C \subseteq N_l(s)$ through queue $Q$. For each node $v \in C$, the distance $d_{ST}(v, x)$ is calculated for all neighborhoods $x$ of $v$. Using the data structures $S$ and $T$, this takes $O(\log n + A)$. \textbf{SToC} iterates over seeds $s$ by removing from the graph nodes that appear in the cluster $C$ returned by \textbf{STo-Query}. This implies that a node is enqued in $Q$ exactly once. In summary, worst-case complexity of \textbf{SToC} is 
$O(\sum_{x\in V} |N(x)|(\log n + A)) = O(m(\log n + A))$.  According to related work, we consider $A$ to be constant in real world datasets. This leads to an overall time complexity of $O(m \log n)$. The initialization of the data structures $S$ and $T$ has the same cost. In fact, $S$ can be filled in linear time $O(n)$ through a scan of the input attributed graph. Moreover, bottom-k sketches can be computed in $O(m k)$ time \cite{boldi2011hyperanf}, hence, for $k \in O(\log{n})$, building $T$ requires $O(m \log n)$.

Regarding space usage, the dictionary $S$ requires $O(nA) = O(n)$ space, assuming $A$ constant. Moreover, since each sketch vector in $T$ has size $O(\log{n})$, $T$ requires $O(n\log{n})$ space. Thus, \textbf{SToC} requires $O(n \log{n})$ space, in addition to the one for storing the input graph.

\section{Auto-tuning of parameters}\label{sec:tuning}

The \textbf{SToC} algorithm assumes two user parameters\footnote{The  error threshold $\epsilon$ is more related to implementation performance issues rather than to user settings.} in input: the value $l$ to be used in topological distance (Equ.~\ref{eq:structuralDistance}), and the distance threshold $\tau$ tested at line 7 of  \textbf{STo-Query}. The correct choice of such parameters can be critical and non-trivial. For example, consider the cumulative distributions of $d_S()$ and $d_T()$ shown in Fig.~\ref{fig:ddists} for the datasets that will be considered in the experiments. Small values of $l$  make most of the pairs very distant, and, conversely, high values of $l$ make most of the pairs close w.r.t.~topological distance. Analogously, high values of threshold $\tau$ may lead to cluster together all nodes, which have semantic and topological distance both lower than $\tau$. E.g.,~almost every pair of nodes has a semantic distance lower than $0.4$ for the DIRECTORS dataset in Fig.~\ref{fig:ddists}.

Another issue with parameters $l$ and $\tau$ is that they are operational notions, with no clear intuition on how they impact on the results of the clustering problem. In this section, we introduce a declarative notion, with a clear intuitive meaning for the user, and that allows to derive optimal values for $l$ and $\tau$. 
We define the \emph{attraction ratio} $\arat$ as a value between 0 and 1, as a specification of the expected fraction of nodes similar to a given one.  Extreme values $\arat = 1$ or $\arat = 0$ mean that all nodes are similar to each other and all nodes are different from each other respectively. In order to let the user weight separately the semantic and the topological components, we actually assume that a semantic attraction ratio $\arat_S$ and a topological attraction ratio $\arat_T$ are provided by the user. We describe next how the operational parameters $l$ and $\tau$ are computed  from the declarative ones $\arat_S$ and $\arat_T$.

\emph{Computing $\tau$.}
We approximate the cumulative distribution of $d_S()$ among all the $n^2$ pairs of nodes by looking at a sample of  $\frac{2 \log n}{\epsilon^2}$ pairs. By the Hoeffding bound~\cite{crescenzi2011comparison}, this guarantees error $\epsilon$ of the approximation. Then, we set $\tau = \hat{\tau}$ as the $\arat_S$-quantile of the approximated distribution. By definition, $d_S()$ will be lower or equal than $\hat{\tau}$ for the fraction  $\arat_S$ of pairs of nodes. Fig.~\ref{fig:ddists} (second and fourth plots) show the approximated distributions of $d_S()$ for the DIRECTORS and DBLP datasets. E.g.,~an attraction ratio $\alpha_S = 0.4$ can be reached by choosing $\tau = 0.2$  for DIRECTORS and $\tau = 0.45$ for DBLP.

\begin{figure}
    \centering
    \includegraphics[width=0.9\columnwidth]{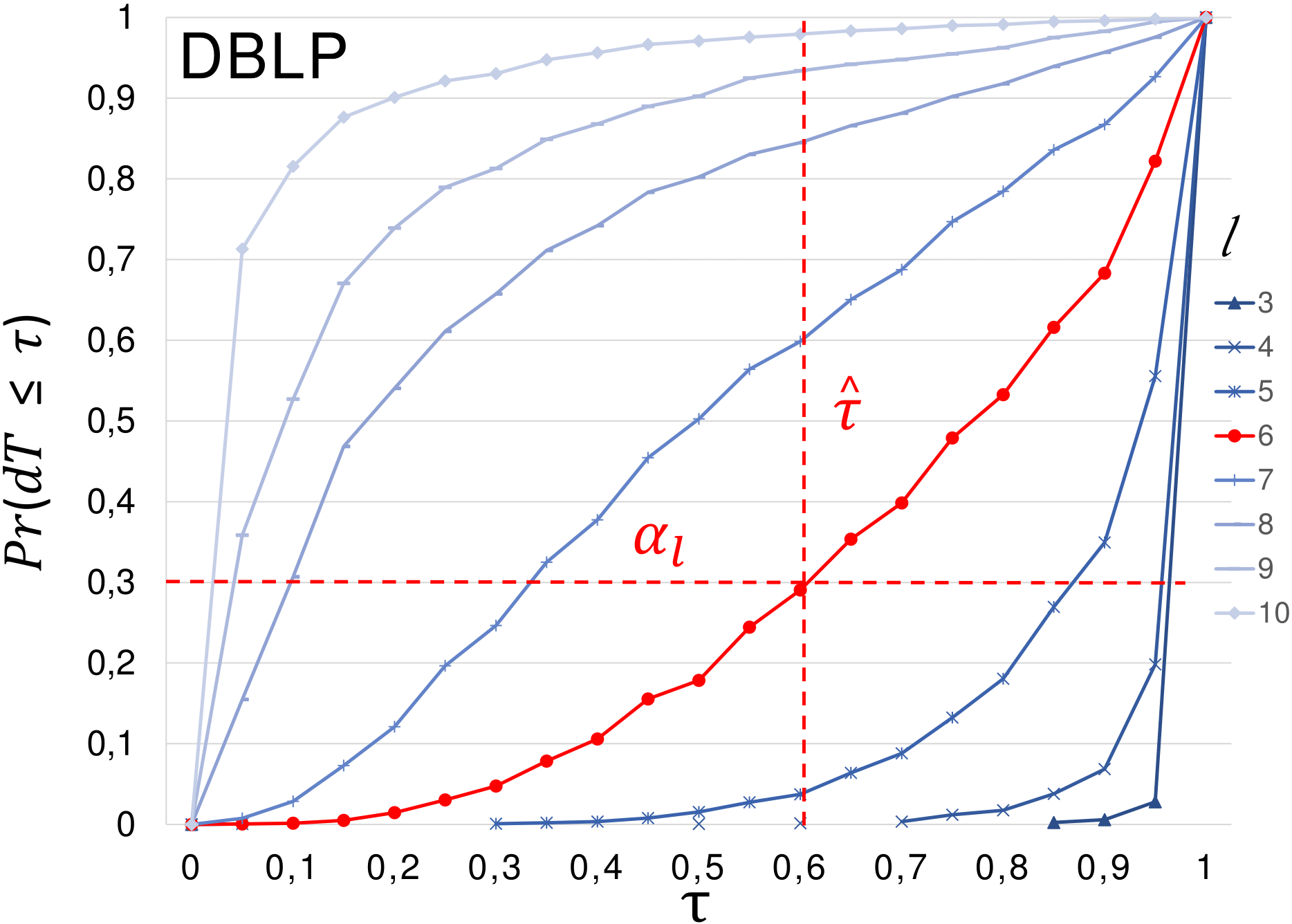}
    \caption{Computing the $l$ value for $\arat_T=0.3$ and $\tau=0.6$ on  the DBLP dataset.}
    \label{fig:radius_lines}
\end{figure}

\emph{Computing $l$.} In the previous step, we fixed $\tau = \hat{\tau}$ using the semantic distance distribution. We now fix $l$ using the topological distance distribution. The approach consists in approximating the cumulative distribution of $d_T()$, as done before, for increasing values of $l$ starting from $l=1$. For each value of $l$, we look at the quantile of $\hat{\tau}$, namely the fraction $\alpha_l = Pr( d_T \leq \hat{\tau})$ of pairs of nodes having topological distance at most $\hat{\tau}$. We choose the value $l$ for which $|\alpha_l - \alpha_T|$ is minimal, namely for which $\alpha_l$ is the closest one to the attraction ratio $\alpha_T$. Since $\alpha_l$ is monotonically decreasing with $l$, we stop when  $|\alpha_{l+1} - \alpha_T| > |\alpha_l - \alpha_T|$. Fig.~\ref{fig:radius_lines} shows an example for $\arat_T=0.3$ and $\hat{\tau}=0.6$. The value $l=6$ yields the quantile $\alpha_l$ closest to the expected attraction ratio $\arat_T=0.3$.

We now show that the cost of the auto-tuning phase is bounded by $O(m\log n)$, under the assumptions that both $\epsilon^{-1}$ and $l$ are $O(1)$. Such assumption are realistic. In fact, values for $\epsilon$ cannot be too small, otherwise the performance improvement of using bottom-$k$ sketches would be lost~\cite{cohen2007summarizing}. Regarding $l$, it is bounded by the graph diameter, which for real-world networks, can be considered bounded by a constant~\cite{backstrom2012four,ugander2011anatomy}. 
Let us consider then the computational cost of auto-tuning. 
Computing $\tau$ requires calculating semantic distance among $\frac{2 \log n}{\epsilon^2}$ pairs of nodes, which requires $O(\log^2 n)$, and in sorting the pairs accordingly, which requires $O( (\log n)(\log \log n) )$. Overall, the cost is $O(\log^2 n)$. 
Computing $l$ requires a constant-bounded loop. In each iteration, we need to build an approximate cumulative distribution of the topological distance $d_T()$, which, as shown before, is $O(\log^2 n)$. In order to compute $d_T()$ we also have to construct the data structure $T$ for the value $l$ at each iteration, which requires $O(m\log n)$. In summary, computing $l$ has a computational cost that is in the same order of the \textbf{SToC} algorithm.

\section{Experiments}\label{sec:experiments}

We first present the evaluation metrics and the experimental datasets, and then show the results of our experiments, which aim at showing the effectiveness of the approach and at comparing it with the state of the art.

\subsection{Evaluation metrics}

According to existing approaches~\cite{duong2013filtering,xu2014gbagc,zhou2010clustering}
we adopt both a semantic and a topological metric. %
The semantic metric is the \emph{Within-Cluster Sum of Squares} (WCSS) also called \emph{Sum of Squared Error} (SSE), widely used by widespread approaches such as the $k$-means~\cite{lloyd1982least}: %
\begin{equation}
WCSS = \sum_{i=1}^{k} \sum_{ v \in C_i} \left\| v - \mu_i \right\|^2
\label{equ:WCSS}
\end{equation}
where $C_1, \ldots, C_k$ is the clustering of the graph, and, for $i \in [1, k]$, $\mu_i$ is the centroid of nodes in $C_i$ w.r.t.~the semantics distance $d_S()$~\cite{protter1977college}. WCSS ranges over non-negative numbers, with lower values denoting better clusterings. Alternative semantic metrics, such as entropy \cite{xu2014gbagc,zhou2010clustering}, are suitable for categorical/discretized attributes only.

The topological metric is the \emph{modularity}~\cite{clauset2004finding}, a de-facto standard for graph clustering evaluation~\cite{xu2014gbagc,zhou2010clustering}: 
\begin{equation}
Q = \frac{1}{2m} \sum_{v,w\in V} \left[ A_{vw} - \frac{k_v k_w}{2m} \right] \delta (c_v,c_w)
\label{equ:modularity}
\end{equation}
where $A$ is the adjacent matrix of the graph, $k_v$ is the degree of node $v$, $c_v$ is the cluster ID of node $v$, and $\delta$ is the identity function ($\delta(i,j) = 1$ if $i=j$ and 0 otherwise).
$Q$ is defined in $[-\frac{1}{2},1]$, with a random clustering expected to have $Q=0$, and a good clustering $Q>0$~\cite{clauset2004finding}. 
\subsection{Experimental datasets}

We will run experiments on two datasets, whose summaries are shown in Table~\ref{tab:data}.

\emph{DIRECTORS: A social network of directors.} A \emph{director} is a person appointed to serve on the board of a company. 
We had a unique access to a snapshot of all Italian boards of directors stored in the official Italian Business Register. 
The attributed graph is build as follows: nodes are distinct directors, and there is an (undirected) edge between two directors if they both seat in a same board. In other words, the graph is a bipartite projection 
of a bipartite graph directors-companies. In the following we distinguish the whole graph (DIRECTORS) from its giant connected component (DIRECTORS-gcc). 
Node attributes include quantitative characteristics of directors (age, geographic coordinates of birth place and residence place) and categorical characteristics of them (sex, and the set of industry sectors of companies they seats in the boards of -- e.g.,~such as IT, Bank, etc.).
\begin{table}[t]
    \centering
    \scalebox{0.85}{
        \begin{tabular}{|r|r|l|l|}
            \hline
            \multicolumn{2}{|c|}{\rule{0pt}{2ex}\emph{Topology}} & \multicolumn{2}{c|}{\emph{Attributes}}\\ \hline
            \multicolumn{4}{|l|}{\rule{0pt}{2ex}\textbf{DBLP}}\\\hline
\rule{0pt}{2ex} \emph{Nodes} & \emph{Edges} & \emph{Categorical} & \emph{Quantitative} \\   \hline
\rule{0pt}{2ex} 60,977 & 774,162 & topic & prolific \\ \hline
            \multicolumn{4}{|l|}{\rule{0pt}{2ex}\textbf{DIRECTORS}}\\     \hline
\rule{0pt}{2ex} \emph{Nodes} & \emph{Edges} & \emph{Categorical} & \emph{Quantitative} \\   \hline
\rule{0pt}{2ex} 3,609,806 & 12,651,511 & sectors, sex & age, birthplace, residence \\\hline
            \multicolumn{4}{|l|}{\rule{0pt}{2ex}\textbf{DIRECTORS-gcc}}\\ \hline
\rule{0pt}{2ex} \emph{Nodes} & \emph{Edges} & \emph{Categorical} & \emph{Quantitative} \\   \hline
\rule{0pt}{2ex} 1,390,625 & 10,524,042 & sectors, sex & age, birthplace, residence \\\hline
        \end{tabular}
        }
    \vspace{5pt}
    \caption{Summary of experimental datasets.}
    \label{tab:data}
\end{table}
Clustering this network means finding communities of people tied by business relationships. A combined semantic-topological approach may reveal patterns of structural aggregation as well as social segregation~\cite{baroni2015segregation}. For example, clusters may reveal communities of youngster/elderly directors in a specific sub-graph.

\emph{DBLP: Scientific coauthor network.} This dataset consists of  the DBLP bibliography database restricted to four research areas:
databases, data mining, information retrieval, and artificial intelligence. The dataset was kindly provided by the authors of~\cite{zhou2010clustering}, where it is used for evaluation of their algorithm.
Nodes are authors of scientific publications. An edge connect two authors that have co-authored at least one paper. Each node has two attributes: \emph{prolific} (quantitative), counting the number of papers of the author, and \emph{primary topic} (categorical), reporting the most frequent keyword in the author's papers.

Fig.~\ref{fig:ddists} shows the cumulative distributions of semantic and topological distances for the datasets. In particular, the 1\textit{st} and 3\textit{rd} plots show the impact of the $l$ parameter on the topological distance. The smaller (resp. larger) $l$, the more distant (resp. close) are nodes. This is in line with the small-world phenomenon in networks~\cite[Fig.2]{watts1999networks}.

\subsection{Experimental results}

We will compare the following algorithms:
\begin{itemize}
    \item \textbf{Inc-C}: the \emph{Inc-Clustering} algorithm by Zhou et al.~\cite{zhou2010clustering}. It requires in input the number $k$ of clusters to produce. Implementation provided by the authors. 
    \item \textbf{GBAGC}: the \emph{General Bayesian framework for Attributed Graph Clustering} algorithm by Xu et al.~\cite{xu2014gbagc}, which is the best performing approach in the literature. It also takes $k$ as an input. Implementation provided by the authors.
    \item \textbf{SToC}: our proposed algorithm, which takes in input the attraction ratios $\arat_S$ and $\arat_T$, and the error threshold $\epsilon$.
    \item \textbf{ToC}: a variant of \textbf{SToC} which only considers topological information (nodes and edges). It takes in input $\arat_T$ and the error threshold $\epsilon$ ($\tau$ and $l$ are computed as for \textbf{SToC}, with a dummy $\arat_S = \arat_T$).
    \item \textbf{SC}: a variant of \textbf{SToC} which only considers semantic information (attributes). It takes in input $\arat_S$ and the error threshold $\epsilon$.
\end{itemize}

\begin{table}
    \centering
    \scalebox{0.9}{
    \begin{tabular}{    |p{0.05\columnwidth}
                        |p{0.15\columnwidth}
                        |p{0.13\columnwidth}
                        |p{0.12\columnwidth}
                        |p{0.12\columnwidth}
                        |p{0.1\columnwidth}|
    }
        \hline
        \rule{0pt}{2ex}
        $\arat$& $k$ & Q&$WCSS$&Time (s)&Space (GB)\\
        
        \hline
        \multicolumn{6}{|l|}{\rule{0pt}{2ex}\textbf{DBLP}}\\
        \hline
        \rule{0pt}{2ex} 0.1 &25,598 (1,279) & 0.269 (0.014) & 5,428 (520)   & 0.627 (0.21) & 0.65\\\rule{0pt}{2ex}
        0.2 &26,724 (1,160) & 0.270 (0.015) & 5,367 (534)   & 0.616 (0.214)& 0.64\\\rule{0pt}{2ex}
        0.4 &6,634 (5,564)& 0.189 (0.12)    & 23,477 (3,998)& 0.272 (0.151)& 0.67\\\rule{0pt}{2ex}
        0.6 &9,041 (7,144)& 0.153 (0.08)    & 21,337 (5,839)& 0.281 (0.1)  & 0.7 \\\rule{0pt}{2ex}
        0.8 &11,050 (4,331) & 0.238 (0.1)   & 20,669 (3,102)& 0.267 (0.11) & 0.69\\\rule{0pt}{2ex}
        0.9 &15,041 (5,415) & 0.188 (0.045) & 16,688 (4,986)& 0.28 (0.05)  & 0.64\\
        \hline
        \multicolumn{6}{|l|}{\rule{0pt}{2ex}\textbf{DIRECTORS}}\\
        \hline
        0.1&2,591,184 (311,927)&0.1347 (0.0237)&7,198 (5,382)&3,698 (485)& 9 \\
        0.2&2,345,680 (162,444)&0.1530 (0.0084)&9,793 (1,977)&3,213 (167.3)&8.2\\
        0.4& 1,891,075 (34,276)& 0.22 (0.023)&26,093 (1,829)&2,629 (178.6)&8.6\\
       0.6& 1,212,440 (442,011) & 0.28 (0.068) & 72,769 (4,129) & 17,443 (2,093) &8.7\\
        0.8& 682,800 (485,472)& 0.1769 (0.066)&49,879 (8,581)& 8,209 (12,822)&9.5\\
        \hline
        \multicolumn{6}{|l|}{\rule{0pt}{2ex}\textbf{DIRECTORS-gcc}}\\
        \hline
        0.1&886,427 (123,420) & 0.102 (0.016)&6158 (2886)&278.4 (44.7)&10\\\rule{0pt}{2ex}
        0.2&901,306 (47,486) & 0.103 (0.02)&4773 (2450)&274 (14.1)&10.4\\\rule{0pt}{2ex}
        0.4&811,152 (28,276) &0.1257 (0.0164) &8,050 (1450)&239.1 (11.6)&4.3\\\rule{0pt}{2ex}
        0.6&664,882 (63,334) & 0.248 (0.0578)&13,555 (3,786)&181.6 (24.4)&4.2\\\rule{0pt}{2ex}
        0.8& 584,739 (408,725)& 0.189 (0.0711)&49,603 (9,743) &5,979 (10,518)&8.8\\\hline
    \end{tabular}
    }
    \vspace{5pt}
    \caption{\textbf{SToC} results (mean values over 10 runs, StDev in brackets).}
    \label{tab:SToC}
\end{table}
\begin{table}[t]
    \centering
    \scalebox{0.9}{
        \begin{tabular}{|r|r|r|r|r|}
        \hline
        $k$ & Q & $WCSS$ & Time (s) & Space (GB)\\
        \hline
       \rule{0pt}{2ex} 15  & -0.499 & 38,009 & 870 & 31.0 \\\rule{0pt}{2ex}
        100     & -0.496  & 37,939 & 1,112  & 31.0 \\\rule{0pt}{2ex}
        1,000   & -0.413  & 37,051 & 1305   & 31.1 \\\rule{0pt}{2ex}
        5,000   & -0.136  & 32,905 & 1,273  & 32.1 \\\rule{0pt}{2ex}
        15,000  & 0.083   & 13,521 & 1,450  & 32.2 \\
        \hline
    \end{tabular}
    }
    \vspace{5pt}
    \caption{\textbf{Inc-C} results for the DBLP dataset.}
    \label{tab:IncC}
\end{table}

\begin{table}
    \centering
    \scalebox{0.9}{
    \begin{tabular}{|r|r|r|r|r|}
        \hline
      \rule{0pt}{2ex}  $k$ (actual) & Q&$WCSS$&Time (s)&Space (GB)\\
      
        \hline
        \multicolumn{5}{|l|}{\rule{0pt}{2ex}\textbf{DBLP}}\\
        \hline
        \rule{0pt}{2ex} 10 (10)    & 0.0382  &27,041 & 17 &0.5 \\\rule{0pt}{2ex}
        50 (14)    & 0.0183  &27,231      & 25 & 0.6  \\\rule{0pt}{2ex}
        100 (2)~    & $1\cdot 10^{-7}$ &27,516 & 13 & 0.7 \\\rule{0pt}{2ex}
        1,000 (3)~  & $6\cdot 10^{-4}$     &27,465 & 37 & 3 \\\rule{0pt}{2ex}
        5,000 (2)~  & $2\cdot 10^{-5}$   &27,498  & 222  &14.2 \\\rule{0pt}{2ex}
        15,000 (1)~ & 0           &27,509 & 663 &50.182 \\
        \hline
        \multicolumn{5}{|l|}{\rule{0pt}{2ex}\textbf{DIRECTORS}}\\
        \hline
       \rule{0pt}{2ex} 10 (8)~ & 0.0305 &198797&18 &4.3 \\\rule{0pt}{2ex}
        50 (10) & 0.0599 &198792&63 &12.4 \\\rule{0pt}{2ex}
        100 (8)~ & 0.1020 &198791&120 &22.4\\\rule{0pt}{2ex}
        500 (5)~ & 0.0921 &198790&8,129 &64.3 \\\rule{0pt}{2ex}
        1000 (--)~ &-&-&-&out of mem\\
        \hline
        \multicolumn{5}{|l|}{\rule{0pt}{2ex}\textbf{DIRECTORS-gcc}}\\
        \hline
       \rule{0pt}{2ex} 10 (8)~ & 0.1095&75103&94&3.02\\\rule{0pt}{2ex}
        50 (14)&0.0563&75101&161&5.47\\\rule{0pt}{2ex}
        100 (15)&0.0534&75101&234&9.34\\\rule{0pt}{2ex}
        500 (5)~ &0.0502&75101&1,238&40.3\\\rule{0pt}{2ex}
        1000 (7)~ &0.0569&75101&3,309&59\\\rule{0pt}{2ex}
        1500 (--)~ &--&--&--&out of mem\\
        \hline
        \end{tabular}
    }
    \vspace{5pt}
    \caption{\textbf{GBAGC} results.}
    \label{tab:GBAGC}
\end{table}

All tests were performed on a machine with two Intel Xeon Processors E5-2695 v3 (35M Cache, 2.30 GHz) and 64 GB of main memory running Ubuntu 14.04.4 LTS. 
\textbf{SToC}, \textbf{ToC} and \textbf{SC} are implemented in Java 1.8, \textbf{Inc-C} and \textbf{GBAGC} are developed in MatLab. 

Table~\ref{tab:SToC} shows the results of \textbf{SToC} for $\arat_S = \arat_T = \alpha$  varying from 0.1 to 0.9, and a fixed $\epsilon = 0.9$. For every dataset and $\alpha$, we report the number of clusters found ($k$), the evaluation metrics $Q$ and $WCSS$, the running time, and the main memory usage. As general comments, we can observe that $k$ and $WCSS$ are inversely proportional to $\alpha$, $Q$ is always non-negative and in good ranges, memory usage is limited, and running times are negligible for DBLP and moderate for DIRECTORS and DIRECTORS-gcc. For every $\alpha$, we executed 10 runs of the algorithm, which uses random seeds, and reported mean value and standard deviation. The low values of the standard deviations of $Q$ and $WCSS$ show that the random choice of the seeds does not impact the stability of the results. The results of \textbf{ToC} and \textbf{SC} can be found in in Appendix~\ref{apx:sctoc}: 
in summary, the exploitation of both semantic and topological information leads to a superior performance of \textbf{SToC} w.r.t.~both $Q$ and $WCSS$.

Tables~\ref{tab:IncC} and~\ref{tab:GBAGC} report the results for \textbf{Inc-C} and \textbf{GBAGC} respectively. Due to the different input parameters of such algorithms,  we can compare the results of \textbf{SToC} only by looking at rows with similar $k$. Let us consider \textbf{Inc-C} first. Running times are extremely high, even for the moderate size dataset DBLP. It was unfeasible to obtain results for DIRECTORS. Space usage is also high, since the algorithm is in $O(n^2)$. Values of $Q$ are considerably worse than \textbf{SToC}. $WCSS$ tends to generally high.
Consider now \textbf{GBAGC}. 
Quality of the results is considerably lower than \textbf{SToC} both w.r.t.~$Q$ and $WCSS$. 
The space usage and elapsed time increase dramatically with $k$, which is non-ideal for large graphs, where a high number of cluster is typically expected.
On our experimental machine, \textbf{GBAGC} reaches a limit with $k = 500$ for the DIRECTORS dataset by requiring 65GB of main memory. Even more critical is the fact that the number of clusters actually returned by \textbf{GBAGC} is only a fraction of the input $k$, e.g.,~it is 1 for $k=15,000$ for the DBLP dataset. The user is not actually controlling the algorithm results through the required input. 

\newcommand{\picwdt}{.49\columnwidth}

\begin{figure}[!t]
    \centering
    \includegraphics[width=\picwdt]{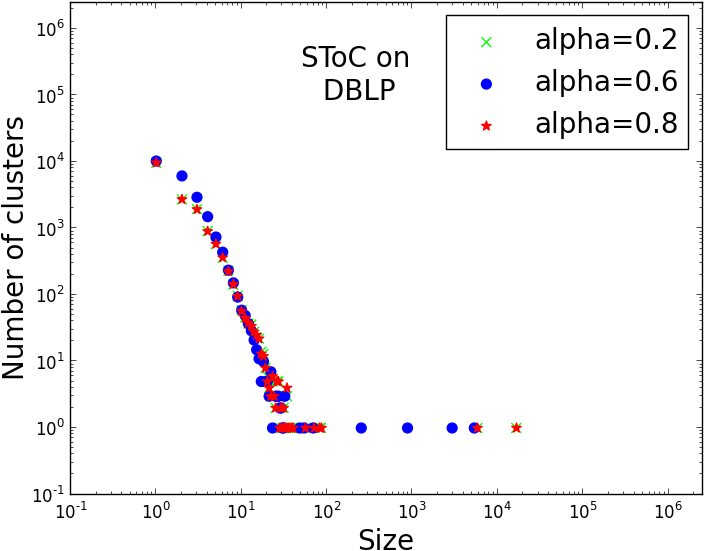} 
    \includegraphics[width=\picwdt]{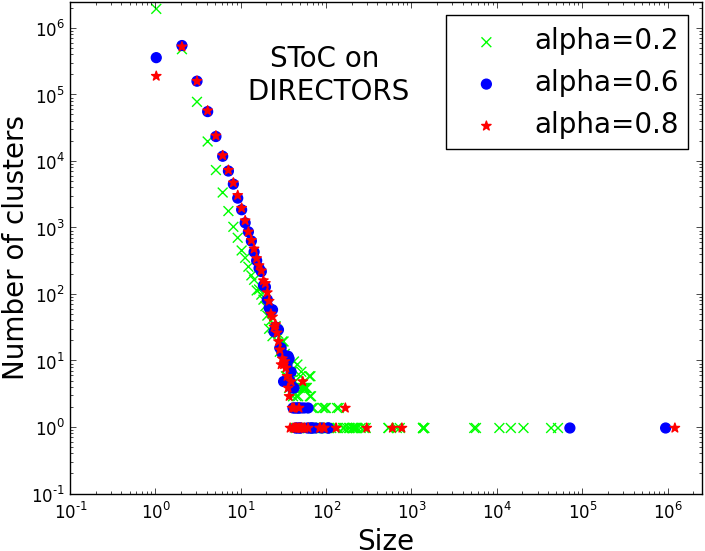}
    \\\vspace*{1pt}
    \includegraphics[width=\picwdt]{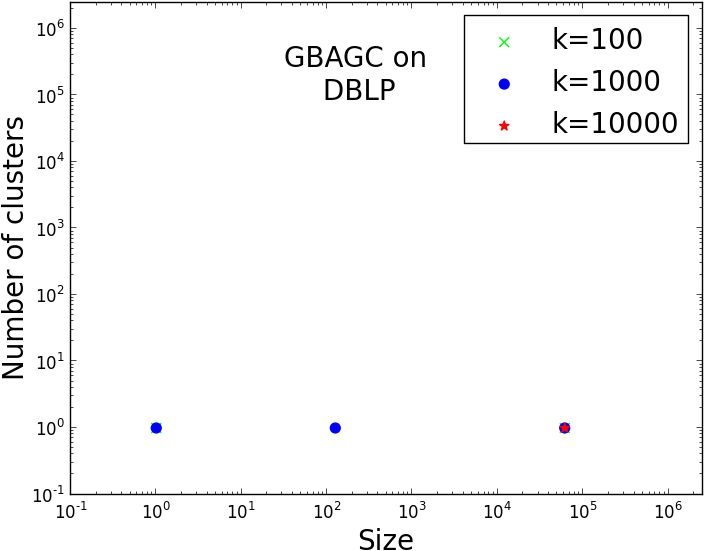} 
    \includegraphics[width=\picwdt]{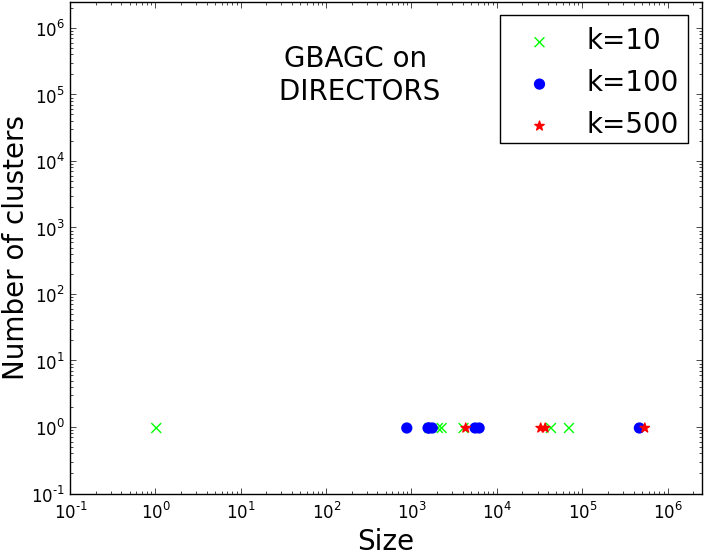} 
    \\\vspace*{1pt}
    \includegraphics[width=\picwdt]{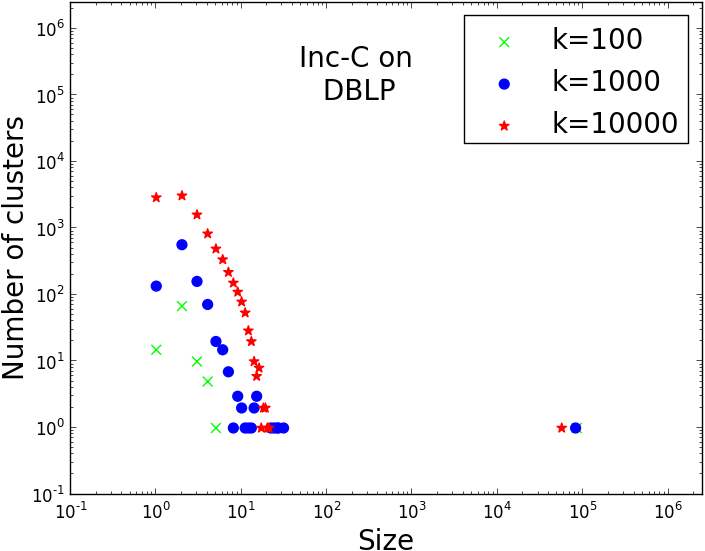}

    \caption{Size distribution of clusters found by \textbf{SToC}, \textbf{GBAGC} and \textbf{Inc-C} for some of the tests in Tables~\ref{tab:SToC}, \ref{tab:IncC}, \ref{tab:GBAGC}.}
    \label{fig:sdist}
\end{figure}
Figure~\ref{fig:sdist} clarifies the main limitation of \textbf{Inc-C} and \textbf{GBAGC} over \textbf{SToC}. It reports for some of the executions the size distributions of clusters found.  \textbf{Inc-C} (bottom plot) tends to produce a single giant cluster including most of the nodes. \textbf{GBAGC} (middle plots) produces a small number of clusters regardless of the input parameters. 
Instead, \textbf{SToC} (top plots) produces more balanced results, typically expected in sociology \cite{bruhn2011sociology,mcmillan1986sense}, with a size distribution in line with common power-laws found in  real-world network and with the input graphs in particular (see \cite{baroni2015segregation} for the DIRECTORS graph).

\section{Conclusions}\label{sec:conclusions}
We proposed \textbf{SToC}, a clustering algorithm for large attributed graphs. It extracts non-overlapping clusters using a combined distance that accounts for network topology and semantic features, based on declarative parameters (attraction ratios) rather than on operational ones (number of clusters) typically required by other approaches.
Experimental results showed that \textbf{SToC} outperforms the state of the art algorithms in both time/space usage and in quality of the clustering found. 

\section*{Acknowledgements}
The authors would like to thank Andrea Marino and Andrea Canciani for useful conversations.

\bibliographystyle{abbrv}
\bibliography{references}

\appendix
\appendices

\section{Validation of the Attribute Model}\label{apx:model}

One of the main differences of the proposed approach with respect to the state-of-the-art is that quantitative attributes do not need to be discretized. We validate the effectiveness of this choice by showing the loss of quality induced by the discretization. Namely, we run the SToC algorithm on the DBLP dataset, where the quantitative attribute represent how prolific an author is, treating the attribute as a categorical one. Table~\ref{tab:discr} shows the results of this process, which can be compared to the results obtained when the quantitative attribute is addressed properly, shown in Table~\ref{tab:SToC}.
We can see not only how both metrics (Q and WCSS) are significantly better when categorical attributes are considered as such, but that ignoring the similarity between similar attributes may lead to an insignificant result, likely due to the flattening of the distances between nodes. This suggests that approaches that handle quantitative attributes may have an inherent advantage with respect to those that need to discretize them.

\begin{table}[!htbp]
    \centering
    \scalebox{0.85}{
    \begin{tabular}{    |p{0.05\columnwidth}
                        |p{0.13\columnwidth}
                        |p{0.15\columnwidth}
                        |p{0.10\columnwidth}
                        |p{0.12\columnwidth}
                        |p{0.1\columnwidth}|
    }
        \hline
        \rule{0pt}{3ex}
        $\arat$& $k$ & Q&$WCSS$&Time (s)&Space (GB)\\
        \hline
        \rule{0pt}{3ex} 0.1 &25,598 (1,279) & 0.269 (0.014) & 5,428 (520)   & 0.627 (0.21) & 0.65\\\rule{0pt}{3ex}
        0.2 &26,724 (1,160) & 0.270 (0.015) & 5,367 (534)   & 0.616 (0.214)& 0.64\\\rule{0pt}{3ex}
        0.4 &6,634 (5,564)& 0.189 (0.12)    & 23,477 (3,998)& 0.272 (0.151)& 0.67\\\rule{0pt}{3ex}
        0.6 &9,041 (7,144)& 0.153 (0.08)    & 21,337 (5,839)& 0.281 (0.1)  & 0.7 \\\rule{0pt}{3ex}
        0.8 &11,050 (4,331) & 0.238 (0.1)   & 20,669 (3,102)& 0.267 (0.11) & 0.69\\\rule{0pt}{3ex}
        0.9 &15,041 (5,415) & 0.188 (0.045) & 16,688 (4,986)& 0.28 (0.05)  & 0.64\\
        \hline
    \end{tabular}
    }
    \vspace{1pt}
    \caption{SToC on the DBLP dataset (average of 10 runs, StDev in brackets).}
    \label{tab:SToC}
\end{table}

\begin{table}[!htbp]
    \centering
        \begin{tabular}{|p{0.05\columnwidth}
                        |p{0.2\columnwidth}
                        |p{0.2\columnwidth}
                        |p{0.21\columnwidth}|
        }
        \hline
        \rule{0pt}{3ex}
        $\alpha$ &$k$ & Q & $WCSS$\\
        \hline
           \rule{0pt}{3ex} 0.1 &25,358 (900) & 0.278 (0.004) & 5,221 (202)\\\rule{0pt}{3ex}
            0.2 &26,340 (1,492) & 0.279 (0.01) & 5,941 (572)\\\rule{0pt}{3ex}
            0.4 &2 (0)& 0.0004 (0) & 60,970 (2.98)\\\rule{0pt}{3ex}
            0.6 &2 (0)& 0.0004 (0) & 60,973 (1.73)\\\rule{0pt}{3ex}
            0.8 &2 (0)& 0.0004 (0) & 60,972 (2.95)\\\rule{0pt}{3ex}
            0.9 &2 (0)& 0.0004 (0) & 60,974 (0.93)\\
            \hline
        \end{tabular}
    \vspace{1pt}
    \caption{A variant of SToC that treats all attributes as categorical on the DBLP dataset (averages of 10 runs, StDev in brackets).}
    \label{tab:discr}
\end{table}

\section{SC and ToC compared to SToC}\label{apx:sctoc}

Table~\ref{tab:nnn} shows, for varying $\arat$, the number of clusters $k$ produced, and the modularity and WCSS of the clustering found by the three variations of our algorithm. The best values for modularity and WCSS are marked in bold.
As one could expect, topological-only algorithm ToC performs poorly w.r.t.~semantic metrics compared to SC and SToC, although semantic-only algorithm SC is competitive with ToC on topology. The clear winner among the three is SToC, which gives a superior performance compared to ToC and SC for most values of $\arat$.
This shows how SToC can effectively combine semantic and topological
information to provide a better clustering.
Table~\ref{tab:nnn} also shows that the number $k$ of clusters in output is inversely proportional to $\alpha$ when topology plays a role, i.e.,~for ToC and SToC. While it is not clear why SC does not seem to follow this behaviour, we suspect it may be due to a small amount of possible $d_S$ values in the DBLP dataset (see Figure~\ref{fig:ddists} in the paper, right).
It is worth noting that the WCSS metric degenerates for high values of $\arat$; this might be due to $\arat \cdot n$ approaching the size of the graph, making any given pair of nodes be considered similar. SC seems more resistant to this degeneration.

\begin{table*}[b]
    \centering
    \begin{tabular}{|l|r|r|r|l|l|l|r|r|r|r|r|}
        \hline
        \rule{0pt}{3ex}&  \multicolumn{3}{c|}{k}  &\multicolumn{3}{c|}{Q}&\multicolumn{3}{c|}{$WCSS$}\\
        \cline{2-10}
        \rule{0pt}{3ex}
        $\alpha$ & SToC & SC & ToC & SToC & SC & ToC & SToC & SC & ToC\\
        \hline
     \rule{0pt}{3ex}
        0.1 & 25,598 & 15 & 1430 & \textbf{0.269} & ~0.0116 &  ~0.0457 & \textbf{5,428} & 14,699 & 19,723\\
        0.2 & 26,724 & 18 & 7574 & \textbf{0.270} & ~0.01554  & -0.03373 & \textbf{5,367} & 14,891 & 16,634\\
        0.4 & 6,634 & 15 & 1420 & \textbf{0.189} & -0.00256 & ~0.04555 & 23,477 & \textbf{14,453} & 19,377\\
        0.6 & 9.041 & 17 & 1174 & \textbf{0.153} & ~0.01248 & -0.1582 & 21,337 & \textbf{14,944} & 20,713\\
        0.8 & 11,050 & 16 & 314 & \textbf{0.238} & -0.00356 &  -0.3733 & 20,669 & \textbf{15,077} & 26,342\\
        0.9 & 15,041 & 16 & 1 & \textbf{0.188} & ~0.00378 & -0.4831 & 16,688 &  \textbf{15,073} & 27,505\\
        \hline
    \end{tabular}
    \vspace{1pt}
    \caption{SToC, SC and ToC on the DBLP dataset}
    \label{tab:nnn}
\end{table*}

\end{document}